\documentclass[fleqn,10pt]{wlscirep}
\usepackage[utf8]{inputenc}
\usepackage[T1]{fontenc}
\usepackage{xr}
\usepackage{comment}

\newif\ifsubmit

\submittrue
\ifsubmit

    \newcommand{\johnpaul}[1]{}
    \newcommand{\cat}[1]{}
    \newcommand{\can}[1]{}
    \newcommand{\giacomo}[1]{}
    \newcommand{\todo}[1]{}
    \newcommand{\tocite}[1]{}
    
    \newcommand{\deletion}[1]{}

\else
    
    \definecolor{comments}{rgb}{0.3, 0.7, 1.0}
    \newcommand{\johnpaul}[1]{[{\color{comments}John Paul: #1}]}
    \newcommand{\cat}[1]{[{\color{comments}Cat: #1}]}
    \newcommand{\can}[1]{[{\color{comments}Can: #1}]}
    \newcommand{\giacomo}[1]{[{\color{comments}Giacomo: #1}]}
    \newcommand{\todo}[1]{[{\color{red}TODO: #1}]}
    \newcommand{\tocite}[1]{[{\color{red}CITE: #1}]}
    
    \newcommand{\deletion}[1]{}

\fi

\makeatletter
\newcommand*{\addFileDependency}[1]{
  \typeout{(#1)}
  \@addtofilelist{#1}
  \IfFileExists{#1}{}{\typeout{No file #1.}}
}
\makeatother

\newcommand*{\myexternaldocument}[1]{%
    \externaldocument{#1}%
    \addFileDependency{#1.tex}%
    \addFileDependency{#1.aux}%
}
\myexternaldocument{supplementary}

\title{Tree-based machine learning performed in-memory with memristive analog CAM}

\author[1,*]{Giacomo Pedretti}
\author[1,*]{Catherine E. Graves}
\author[1,2]{Can Li}
\author[1]{Sergey Serebryakov}
\author[1]{Xia Sheng}
\author[1]{Martin Foltin}
\author[2]{Ruibin Mao}
\author[1,*]{John Paul Strachan}
\affil[1]{Hewlett Packard Labs, Milpitas (CA), USA}
\affil[2]{Hong Kong University, Hong Kong SAR, China}
\affil[*]{giacomo.pedretti@hpe.com, catherine.graves@hpe.com, john-paul.strachan@hpe.com}

\begin{abstract}
Tree-based machine learning techniques, such as Decision Trees and Random Forests, are top performers in several domains as they do well with limited training datasets and offer improved interpretability compared to Deep Neural Networks (DNN). However, while  easier to train, they are difficult to optimize for fast inference without accuracy loss in von Neumann architectures due to non-uniform memory access patterns. Recently, we proposed a novel analog, or multi-bit, content addressable memory (CAM) for fast look-up table operations. Here, we propose a design utilizing this as a computational primitive for rapid tree-based inference. Large random forest models are mapped to arrays of analog CAMs coupled to traditional analog random access memory (RAM), and the unique features of the analog CAM enable compression and high performance. An optimized architecture is compared with previously proposed tree-based model accelerators, showing improvements in energy to decision by orders of magnitude for common image classification tasks. The results demonstrate the potential for non-volatile analog CAM hardware in accelerating large tree-based machine learning models.
\end{abstract}
\begin{document}
\flushbottom
\maketitle
\thispagestyle{empty}

Deep neural networks (DNN) are becoming the mainstream model for numerous classification tasks such as image and voice recognition \cite{lecun_deep_2015}. However, DNN impact is limited for a range of applications where inspectability and explainability are critical, training data may be limited, and/or where domain knowledge and historical expertise needs to be incorporated in critical decisions \cite{gunning_explainable_2017,kaggle_state_2020}. Examples include cybersecurity systems \cite{xin_machine_2018}, decision making and situational intelligence with multimodal inputs (medical, autonomous driving systems)\cite{janai_computer_2020}, and global system modelling (ground water models, smart factories, robotic fleet deployments) for risk/failure assessments and predictive maintenance \cite{susto_machine_2015}. In these domains, tree-based ensemble methods, such as random forests (RF) \cite{biau_random_2016}, are popular machine learning (ML) approaches for the ease of training, good performance with small data sets \cite{zhou_deep_2017} and reasonable interpretability for domain experts to verify and understand \cite{lundberg_local_2020}. However, large scale tree-based models are difficult to optimize for fast runtime without accuracy loss in von Neumann architectures \cite{tracy_towards_2016}. In von Neumann architectures, memory and computing units are physically separated \cite{von_neumann_first_1945}, which results in high energy consumption and time spent for making data travelling from the processor to memory and back \cite{zidan_future_2018}. Moreover, memory access patterns to the model and feature vector for each decision tree (DT) node are non-uniform, and higher accuracy models require more and deeper DTs, resulting in unpredictable traversal times. State of the art implementations run in super-linear time with DT depth, limiting scalability. Various approaches for speeding up RF \cite{chen_visual_2012,asadi_runtime_2014,lee_vocabulary_2015,tracy_towards_2016} showed mainly incremental improvements as the data-locality access pattern problem remains.\\ 
A new class of accelerators where computation is performed inside the memory, termed in-memory computing (IMC) \cite{ielmini_-memory_2018}, is gaining momentum and accelerators for different applications such as neural networks traning/inference \cite{hu_dot-product_2016,li_efficient_2018,ambrogio_equivalent-accuracy_2018,yao_fully_2020}, image processing\cite{sheridan_sparse_2017,li_analogue_2018} and scientific computing \cite{zidan_general_2018,le_gallo_mixed-precision_2018,sun_solving_2019} have demonstrated performance improvements. A RF IMC accelerator based on complementary metal oxide semiconductor (CMOS) static random access memories (SRAM) has also been proposed \cite{kang_194-njdecision_2018}, but high throughput/low energy operation remains a challenge. In fact, many advantages of IMC approaches are found where a novel class of non-volatile memory (NVM) devices, often dubbed memristors \cite{strukov_missing_2008}, which can operate at low power and high speed \cite{ielmini_resistive_2016}, is used in analog mode by programming algebraic-like problems in a memristor crossbar array \cite{ielmini_-memory_2018}. A different class of IMC primitives include content addressable memories (CAM) \cite{pagiamtzis_content-addressable_2006}, which are traditionally based on SRAM and show good throughput performance at the cost of large power consumption. Memristor-based CAM and ternary CAM (TCAM) circuits have been proposed \cite{guo_resistive_2011,guo_ac-dimm_2013,huang_reram-based_2014,lin_74_2016,graves_memory_2020}, and their ability of performing high performance computation, such as finite automata inference \cite{graves_hybrid_2016} has been unveiled. Given that DT and RF consist of a series of logic AND operations of decision nodes, they can be mapped to CAM \cite{tracy_towards_2016}. In addition, we recently developed an analog CAM based on memristor\cite{li_analog_2020}, which leverages the continuous states to enable multi-bit pattern matching on a single cell.

Here we propose a RF accelerator based on an analog CAM array for tree traversal evaluation coupled with traditional 1-transistor-1-resistor (1T1R) resistive random access memory (RRAM)\cite{li_cmos-integrated_2020} for majority voting, to increase the throughput and decrease the energy consumption in large scale RF inference. First, we present in detail the analog CAM circuit and accurate behavioral model from post-layout simulations. Then, after presenting the mapping techniques for DT and RF, we evaluate performance on an architectural level demonstrating feasibility of the concept and increased performance compared with the state of the art. Our in-memory circuit outperforms existing accelerators throughput by 1000x with a 100x reduction of energy to decision.

\section*{Results}
\subsection*{Analog CAM compact model}

\begin{figure}
\centering
\includegraphics[width=\linewidth]{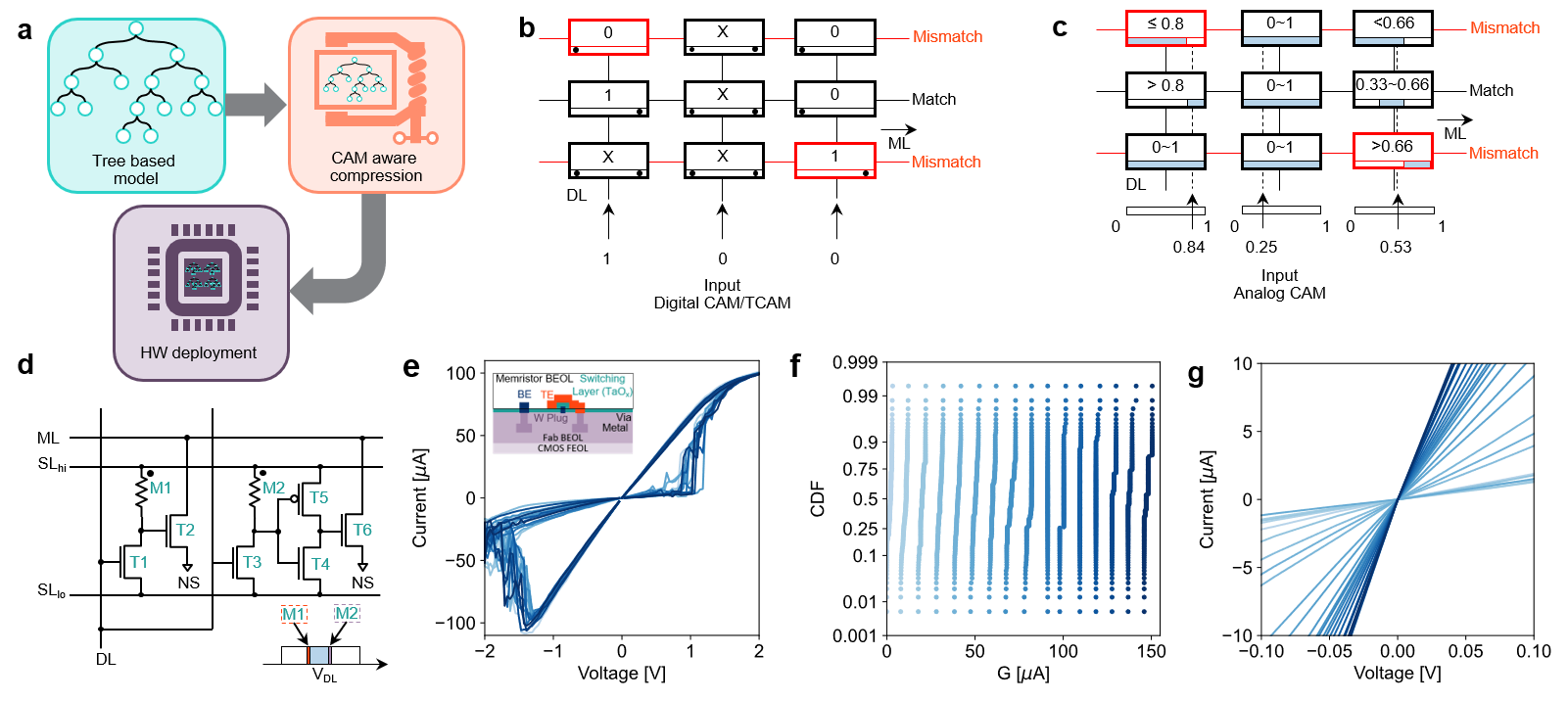}
\caption{Analog content addressable memory with memristor.(a) Illustration of this work, tree-based machine learning models are optimized and deployed on analog CAM hardware. (b) Digital ternary content addressable memory (TCAM) which searches a given input word across the whole memory and outputs the location of a match. (c) Analog CAM, where each cell stores ranges of values, or multi-bit representations. An analog input word is searched against the whole memory array in parallel, similar to the digital TCAM. (d) Circuit schematic of an analog CAM with memristor. (e) Memristor current-voltage (I-V) characteristics for different cycles with corresponding device structure (inset).
(f) Cumulative density functions of 16 levels of conductance corresponding to 4 bits of information, each programmed into 256 memristive devices. 
(g) I-V plot of read sweep for different conductances programmed in memristors showing linear behavior.}
\label{fig:1}
\end{figure}
Tree-based machine learning models can be accelerated by IMC thanks to CAM, a novel class of hardware. Fig. \ref{fig:1}a shows the conceptual flowchart for implementing such model in CAMs, where after the generation of the model it can be compressed before deployment to optimize the performance. Figure \ref{fig:1}b shows the working principle of a digital TCAM. Digital words are stored in different rows of the memory array. By applying a search word on the data line (DL), or columns, it is possible to rapidly search if the word is present in the memory, in which case the address is returned. Each match line (ML), or row, is initially pre-charged and remains charged only if all elements of the searched word match with that stored word.  A wildcard 'X' representing an 'always match' is also possible to be stored or searched, allowing for a third (ternary) state in this TCAM. Different hardware implementations of TCAM have been proposed both based on traditional CMOS technology \cite{pagiamtzis_content-addressable_2006,karam_emerging_2015} and NVM technology \cite{guo_resistive_2011,guo_ac-dimm_2013,huang_reram-based_2014,lin_74_2016,graves_memory_2020}. To represent more than three levels in a TCAM cell one needs to be able to represent a range, defined by a lower and upper limit. For example, if the data $d = 13$ is to be represented, the memory cell should be able to accept any value $12.5<i<13.5$ where the $0.5$  accounts for the system tolerance, namely half of the least significant bit or $LSB/2$. Fig.\ref{fig:1}c shows a conceptual schematic of an analog CAM array \cite{li_analog_2020}, where ranges are stored in memory, an analog word is given as input on the DL (columns) and the corresponding match is sensed on ML (rows) as a digital signal. In this case the equivalent of a wildcard 'X' corresponds to storing the range thresholds to the maximum acceptable limits, in this case 0 to 1, such that any input will match. Fig.\ref{fig:1}d illustrates a schematic for an analog CAM based on memristor technology \cite{li_analog_2020}, although other implementations, for example based on ferroelectric transistors \cite{ni_ferroelectric_2019} have recently been proposed as well. Range thresholds are stored in the memory conductance M1 (lower threshold) and M2 (upper threshold) as shown in the inset. By applying a DL analog value, a voltage divider between memory device and the series transistors T1 and T3 controls the discharge transistor (T2) on the lower threshold side, or the inverter (T4-T5) on the upper threshold side which controls the upper threshold discharge transistor (T6). 

The lower and the higher bound of the searching range is stored as conductance in the RRAM device in our analog CAM. The RRAM device current-voltage (I-V) characteristics are shown in Fig.\ref{fig:1}e, with the device structure fabricated in a back-end-of-line (BEOL) process illustrated in the inset. A TaOx dielectric layer is sandwiched between metallic top electrode (TE) and bottom electrode (BE) and the structure is realized on a conventional 180nm complementary-metal-oxide-semiconductor (CMOS) process \cite{sheng_lowconductance_2019,li_analog_2020,li_cmos-integrated_2020} (see Methods). A newborn device is typically in a high resistance state due to limited conduction in the dielectric layer. After a forming procedure, oxygen vacancies are reordered such that a conductive path is formed between TE and BE resulting in a low resistance state (LRS). Then by applying a negative reset pulse, the conductive path can be retracted and the device results in a high resistance state (HRS). Conductance can be modulated from LRS to HRS by switching positive and negative voltage. Moreover, switching to different intermediate states can be controlled through a variety of means, including through current compliance ($I_C$) modulation, namely the maximum current flowing into the RRAM device during the set transition controlled with a series transistor, or by means of $V_{stop}$, or the maximum voltage applied during the reset operation \cite{ielmini_resistive_2016,ielmini_device_2020}. Fig.\ref{fig:1}f shows the cumulative distribution function of 16 different levels measured on 2048 devices on a large array \cite{li_cmos-integrated_2020} (Supplementary Information Fig.\ref{sifig:1}), corresponding to 4 bits, demonstrating the possibility of analog and multi-bit capability. If a larger number of bits is needed, multiple cells can be used in parallel, with a bit-slicing technique, similar to what is typically done in crosspoint arrays \cite{shafiee_isaac_2016}. For small applied voltages, memristor devices offer a linear conduction as shown in Fig.\ref{fig:1}f, especially for states close to LRS levels. The linear dependence simplifies interpretation of the voltage divider in the analog CAM circuit.
\begin{figure}
\centering
\includegraphics[width=0.9\linewidth]{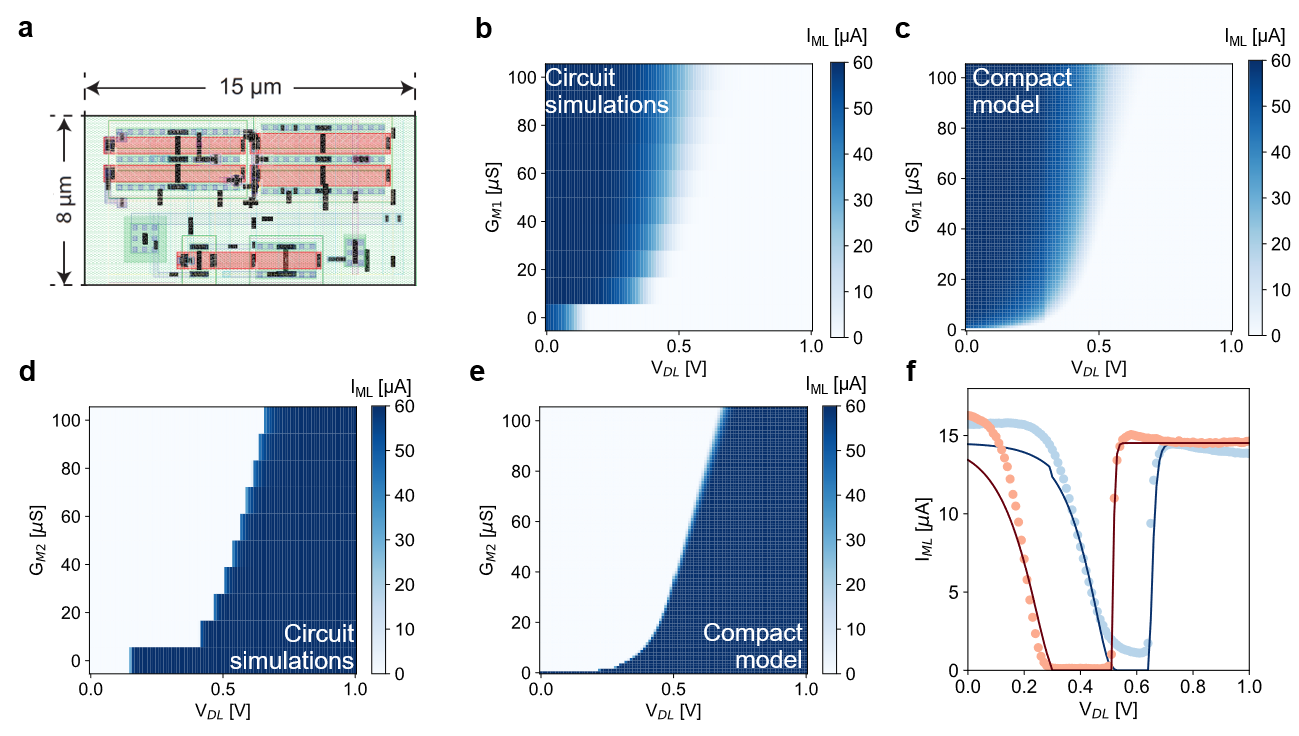}
\caption{Analog CAM compact model. (a) 180 nm cell layout showing the various inputs/outputs. (b,c) Circuit simulation and model calculation of ML discharge current on the lower threshold branch as function of $V_{DL}$ and $G_{M1}$. (d,e) Circuit simulation and model calculation of ML discharge current on the upper threshold branch. (f) Experimental data (circle) and compact model (lines) for two different ranges stored in analog CAM cells}
\label{fig:2}
\end{figure}

An analog CAM array was fabricated in 180 nm CMOS technology \cite{li_analog_2020}, with the cell layout shown in Fig. \ref{fig:2}a. We have also carried out extensive circuit simulations on a more aggressive 16 nm technology node \cite{li_analog_2020} to study the impact of power consumption and scalability of the design. However, to have a fast deployment and performance assessment of more complex and large scale problems, such as DT/RF and other tree-based machine learning based on finite automata \cite{tracy_towards_2016}, a compact and reliable analog CAM cell model should be realized. In fact, SPICE like circuit simulation can only be performed to small scale arrays, masking the true advantage of large scale tasks. For this reason we designed a compact cell model whose details are illustrated in Supplementary Information Fig.\ref{sifig:2} and Methods. Fig.\ref{fig:2}b shows circuit simulation results of the current flowing in the lower threshold branch, or in transistor T2, as function of $V_{DL}$ and M1 programmed conductance $G_{M1}$, where a large current corresponds to a low $V_{DL}$ or high $G_{M1}$ as expected. Fig.\ref{fig:2}c shows the model calculation for the same current which is in good agreement with the circuit simulation. Fig.\ref{fig:2}d-e show the circuit simulation and model calculation, respectively, of the current flowing into the upper threshold branch, or in transistor T6, where in this case high current corresponds to high $V_{DL}$ or low $G_{M2}$. Data and calculation are in good agreement, confirming the model reliability. Note that circuit simulations were performed by taking into account post-layout parasitic effects (see Methods), thus the model comprehensively describes the cell behavior. Fig\ref{fig:2}f shows data (circle) and model calculation (lines) of two different ranges programmed in two analog CAM cell \cite{li_analog_2020}, which confirms that the model can accurately predict cell behavior. 

\subsection*{Mapping DT/RF to analog CAM}
\begin{figure}
\centering
\includegraphics[width=\linewidth]{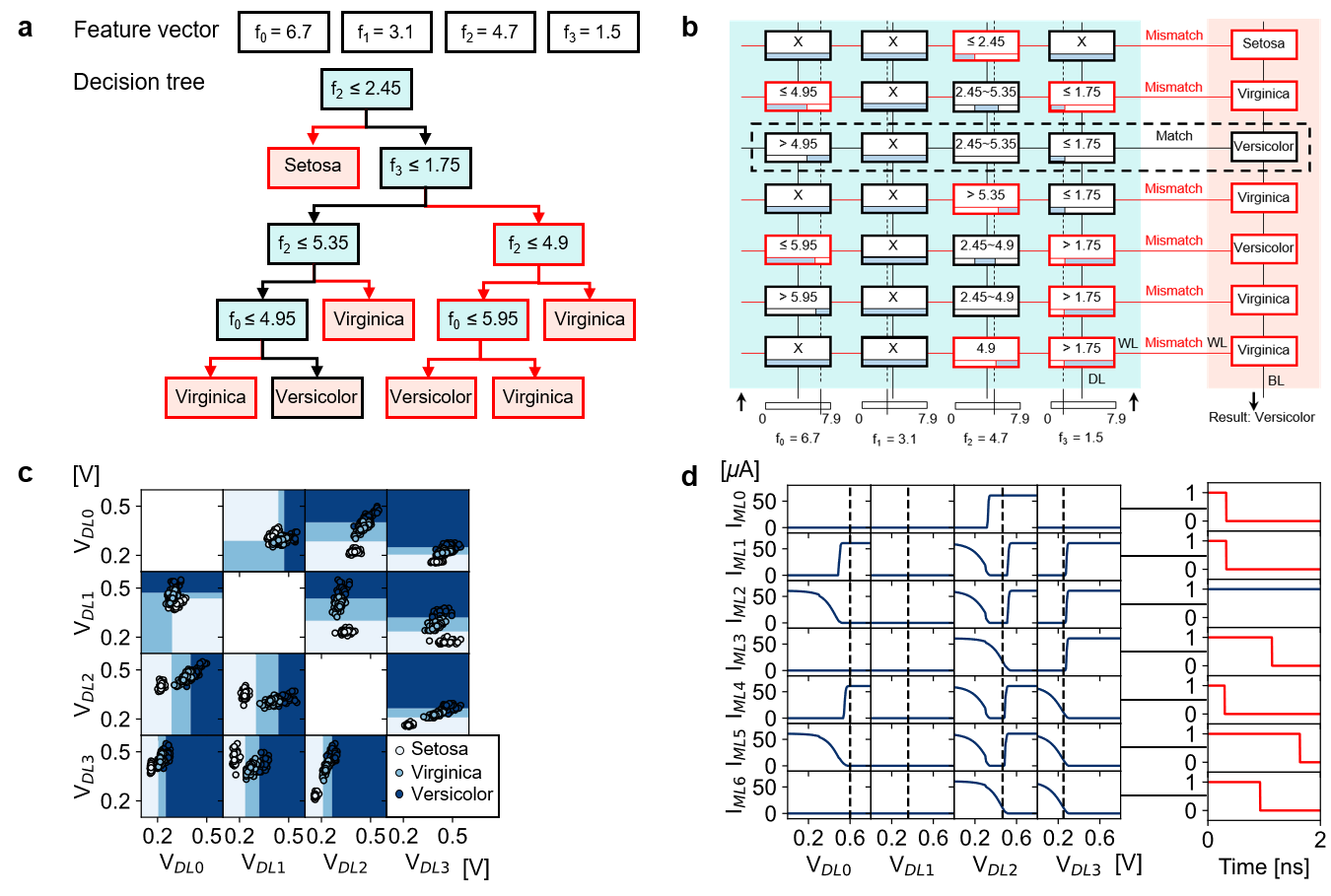}
\caption{Mapping decision tree on analog CAM. (a) Decision tree for classifying the Iris dataset; a feature vector is given as input and features are compared with learned thresholds to decide which branch of the tree to take. (b) Analog CAM array mapping the decision tree of (a), where each root to leaf path is written in an array row. Classification outputs are given on the ML, in one shot. (c) Decision map of different DT trained with two features of the feature vector, calculated in the analog CAM. (d) Current flowing in every anolog CAM cell as function of $V_{DL}$ and resulting ML digital output for the inference of feature vector and DT in (a)
\can{Can we physically program the required conductances (8x7?) in the super-T, and use the experimentally readout conductance in the compact model? If the data shown here is already based on physical conductances, maybe we should show those conductances here.}\giacomo{No the conductance are not based on physical values here. We could do that but it add too much stuff to this figure}
}
\label{fig:3}
\end{figure}
DT and RF are powerful ML models allowing data classification and regression, with much clearer understanding of the resulting models than deep learning techniques. As a toy example, Fig.\ref{fig:3}a shows a DT trained to classify the Iris data set \cite{fisher_use_1936}, where features namely sepal and petal width and length are organized as a feature vector $f = [f_0,f_1,f_2,f_3]$ and given as input. At each node a decision on a feature is made according to a threshold. If the decision is positive, the tree is traversed from top to bottom following the left branch; if the decision is negative, the right branches are taken. Trees are traversed until reaching a leaf, which in this case corresponds to the classes of Iris, namely Setosa, Virginica or Versicolor. DT can be mapped to analog CAM by directly programming each root-to-leaf path into an array row. Feature vectors $f$ are given as input to the columns DL, and ML is initially pre-charged. If all the analog CAM cells of a row match $f$, ML stays charged - otherwise ML discharges into the unmatched analog CAM cell. Note that this corresponds of doing an AND operation between every analog CAM cell in a row. Fig. \ref{fig:3}b shows the implementation of the DT of Fig.\ref{fig:3}a into an analog CAM array. If a feature component is not present in the root-to-leaf path, a wildcard 'X' can be inserted corresponding to the whole range programmed in the analog CAM, i.e. the LRS on the lower threshold memristor and the HRS on the upper threshold memristor, as can be seen in the $f_1$ column. If a feature is present multiple times in a branch and with different thresholds, for example in the second row, third column, then the two threshold are combined and a range is encoded. In the case of only one threshold decision for a particular feature, one of the memristors is kept as wildcard (LRS or HRS) and the other is programmed at an intermediate threshold value, implementing a 'less than or equal to' with a high threshold (a left branch), while a greater than is programmed in the opposite case (a right branch). While the presence of a large number of 'X' appears as a drawback it will be actually used for allowing compression. MLs of the matching rows directly correspond to the matching class, making an analog CAM able to perform a one step classification independent of the array size, corresponding to three clocks cycles $t_{CLK}$ for charging ML, asserting DLs, and latching MLs after $t_{CLK}$. For verifying that analog CAM arrays is effectively able of drawing decision boundaries
we first trained 12 different DTs with all possible intersections of two features, and observe in a two dimensional space the classification results, by deploying the DTs on analog CAM arrays using the compact model and monitoring the matched rows. Fig.\ref{fig:3}c shows a plot of the ML predicted class (shadows) and ground truth results (circles) as a function of different DL voltages corresponding to different feature vector values. Circles landing on a shadow with matching color corresponds to a correct prediction. The trend suggests that the DT model deployed on analog CAM draws the correct decision boundaries in the classification task. Fig.\ref{fig:3}d shows a plot of the ML discharge currents (left) in each analog CAM cell of Fig.\ref{fig:3}b as a function of $V_{DL}$ and the corresponding ML outputs as a function of time for the same feature vector of Fig.\ref{fig:3}a-b, demonstrating the ability to recognize the correct class. The corresponding conductance values mapped in the analog CAM array are shown in Supplementary Information \ref{sifig:3}.
\begin{figure}
\centering
\includegraphics[width=0.8\linewidth]{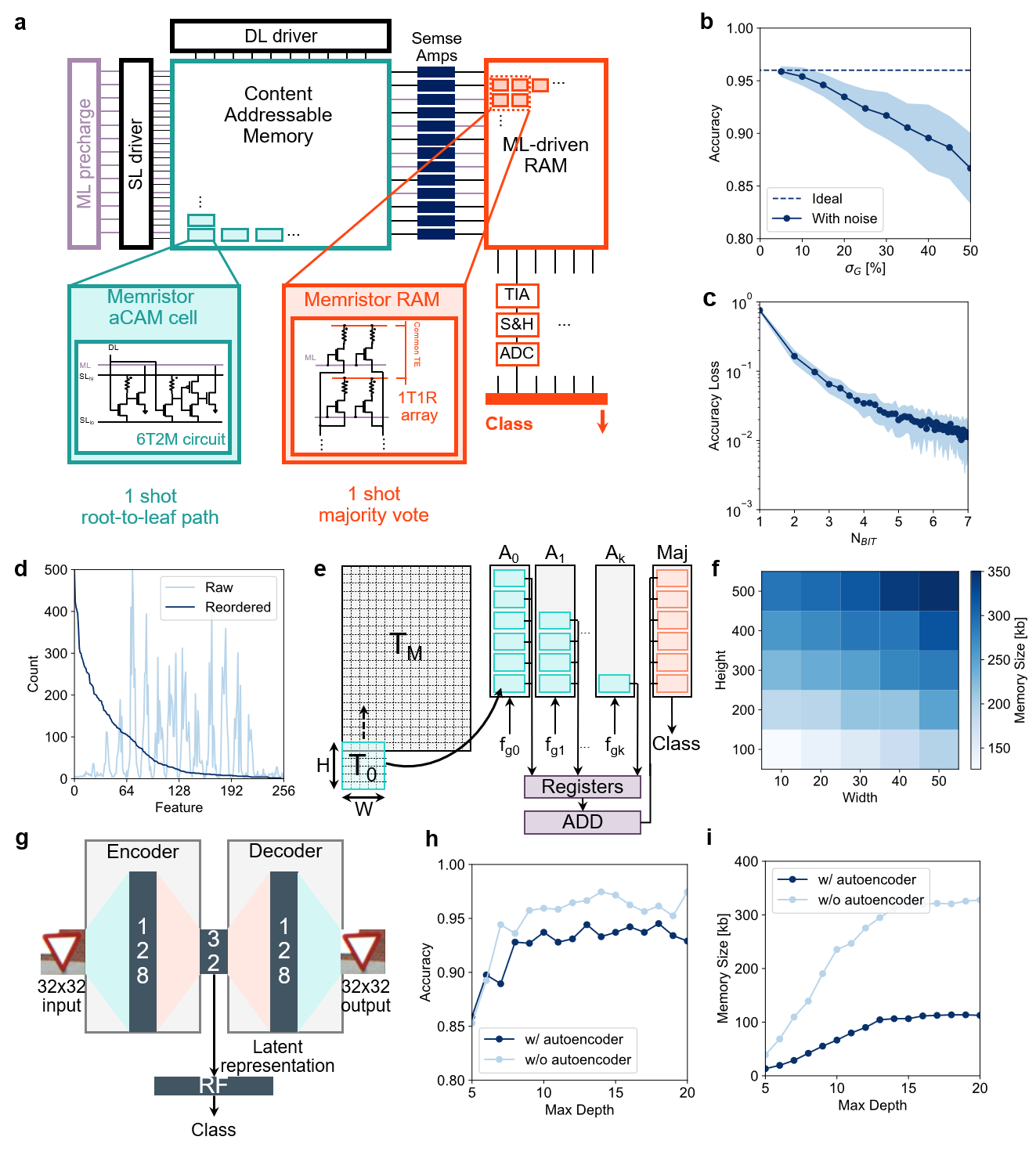}
\caption{RF accelerator architecture. (a) Overview of the full IMC system, with analog CAM executing root-to-leaf evaluation in one step and 1T1R RRAM array executing the majority vote in the analog domain. (b) Accuracy as function of the standard deviation of injected noise $\sigma_G$ in the programmed conductance for 100 different experiments, filled line represent the average while bands corresponds to the standard deviation. (c) Accuracy loss as function of number of bit for representing the threshold in 10 RF trained on the data set, filled line represent the average while band the standard deviation. (d) Number of populated (non `X') analog CAM cells in the arrays as function of the feature index for a direct mapping (raw, light blue line) and reordered array (blue line). (e) Tiles and array mapping procedure, the threshold map TH$_{map}$ traversed from bottom to top and from left to right by $H\times W$ tiles ($T_0$ in the example) which are filled in the presence of valid rows, namely rows that are not completely empty. Once a tile is full, it is placed in the corresponding array ($A_0$) in the example, which will evaluate a group of $W$ features from the feature vector. (f) Memory size needed as function of tile height $H$ and width $W$. (g) Schematic of variational autoencoder architecture for compressing the data set a 
32 wide feature vector. (h) Accuracy as function of maximum depth in the RF with and without autoencoder as input. (i) Memory size required as function of maximum depth in the RF with and without autoencoder as input.}
\label{fig:4}
\end{figure}

While DT are easy to train and deploy, their accuracy for real world problems is affected by over-fitting. This is due to the need for more tree depth to effectively minimize the cost function during training. To avoid this, ensemble methods are used in which multiple trees are evaluated in parallel. This is the case for RF, consisting of $n_{trees}$ inferred in parallel, with the output result computed as majority gate of single DT outputs. Each DT in a forest is trained with a small random portion of the data set and usually requires only a shallow depth and relatively small number $n_{trees}$ to reach good accuracy. Both tree inference and majority vote can be implemented with IMC. Fig. \ref{fig:4}a shows an architectural overview of the RF inference acceleration approach. Each root to leaf path of each DT is mapped to a row of the analog CAM array, whose ML outputs are converted to a digital high or low signal with a sense amplifier. Sense amplifier outputs are connected to the gate of a one-transistor-one-resistor (1T1R) memristor (RRAM) array \cite{li_cmos-integrated_2020}, with every column sensing a corresponding class. The 1T1R RRAM array $M$ is programmed such that $M[i,j]=LRS$ if $class(i)=class(j)$, where $i$ corresponds to the ML index and $j$ the column index, and columns correspond to a different class. In this way the more ML are activated of a given class, the larger the current flowing into the 1T1R RRAM array column for that class. Currents are sensed with a typical chain consisting of a transimpedence amplifier (TIA), sample and hold (S\&H) and analog to digital converter (ADC) \cite{li_cmos-integrated_2020}. Note that only 4 clock cycles, corresponding to pre-charging the ML, asserting DL, evaluating the root to leaf path with the SA latch, and triggering the RRAM read, are needed to reach a classification result and, as a first approximation, this is independent of the number of trees in the forest.\\
To test our system and compare results with previous benchmarks \cite{kang_194-njdecision_2018}, we implemented a RF for the classification of KUL Belgium traffic sign data set \cite{prisacariu_integrating_2010} whose data processing is explained in Supplementary Information \ref{sifig:4}. We mapped the RF into the analog CAM and RRAM arrays and evaluated the accuracy of inference on 200 samples\cite{kang_194-njdecision_2018} reaching 0.965, higher than the reference state of the art. These RF models are well-matched to analog IMC implementations, showing a strong resilience to variation and noise that can otherwise affect analog hardware. As an example, Fig.\ref{fig:4}b shows the RF accuracy as a function of the standard deviation of a Gaussian distribution representing the variability in the memristor conductances, which captures a practical challenge in some memristive devices \cite{ielmini_device_2020}. Accuracy remains unaltered for a standard deviation up to $\sigma_G = 5 \%$, which can be realized in practical and size-scaled devices \cite{sheng_lowconductance_2019}.  Fig. \ref{fig:4}c shows accuracy loss as a function of the number of bits considered in programming the analog CAM threshold demonstrating that only for fairly low bit numbers, i.e. $N_{bit}<3$ the accuracy degrades considerably.

\subsection*{Architecture optimization}
To directly deploy our RF model to an analog CAM, a very large array, i.e. $2000\times256=512$ kb, is needed for one-shot classification. However, most of the analog CAM cells in an RF implementation remain empty. In fact, each root to leaf path has a maximum size of the number of decision nodes, a hyperparameter that can be defined during training and known as maximum depth. Typically the maximum depth is small compared to the number of available features; in the present reference data set the feature vector length is $F=16\times16=256$, and good RF accuracy is reached with a maximum depth $\sim$10 (Supplementary Information Fig.\ref{sifig:5}). Fig.\ref{fig:4}d shows the number of occupied cells in the analog CAM array as a function of the feature identifier. As seen from the 'Raw' (light blue) line, some features occur frequently and correspond to 'important' pixels in the training images (for example defining the shape of the traffic sign), but other features are hardly considered (for example the border of the images) and only a few analog CAM cells of the corresponding columns are occupied. Given that the feature order is arbitrary in each individual root-to-leaf path, we reordered them based on occurrence such that all important features are on the left side of the entire analog CAM array. In this way part of the array remains completely off, or empty. Moreover, we similarly reorder the columns to make sure that the most populated columns are on the bottom of the analog CAM array. The blue line of Fig.\ref{fig:4}d shows the reordered count, evidencing that a part of the array can remain empty and offering compressibility once the large RF array is tiled onto reasonable sized analog CAM arrays. 
We investigated efficient architectures for mapping a large RF model by exploring CAM array tiles sizes and available compression schemes. We divided the CAM architecture into tiles of practical size $H\times W$, i.e. up to $480\times48$, dimensions that were previously found to be feasible \cite{li_analog_2020,graves_memory_2020}. Fig.\ref{fig:4}e shows the tile writing procedure and the tiled architecture. Given a target threshold map TH$_{Map}$ following the reordering procedure we start by sweeping a $H\times W$ tile $T_0$ on the left part of the array, i.e. we evaluate TH$_{Map}[0,0:W]$ and if there is at least one cell not empty we accept the row and write it in $T_0[0,0:W]$, otherwise we discard it. We continue evaluating TH$_{Map}[i,0:W]$, with $i=1$ and writing in $T_0[j,0:W]$ incrementing $i$ at every cycle and $j$ only if the location is written, until $T_0$ is filled. We proceed by positioning $T_0$ at array location $A_0$ which will evaluate the feature group $f_{g0}$ corresponding to the first $W$ features. We take a new tile $T_1$ and start filling it and repeat the process until all elements of TH$_{Map}[:,0:W]$ have been considered. Once this part of TH$_{Map}$ has been mapped, we increment the column, namely we start evaluating TH$_{Map}[0,W:2W]$, and place the corresponding tiles in array location $A_1$, which evaluates feature group $f_{g1}$ corresponding to features $W\sim2W$. The process is repeated until all of TH$_{Map}$ has been evaluated and all the $k=F/W$ arrays are populated. Note that most tiles of $A_0$ are populated while most of $A_k$ are empty, thanks to the reordering. In this way most of the right side arrays tiles can be eliminated.The output of each tile is collected in a register ad logically added to perform the final majority vote only one time in a RRAM memory. Fig. \ref{fig:4}f shows the CAM memory size as a function of the tile dimensions $H$ and $W$ after reordering and mapping, demonstrating significant compression compared to the initial size of $512$ kb. We finally choose the training hyperparameters for benchmarking our system to yield good accuracy with a reduced memory size (Supplementary Information Fig. \ref{sifig:5}), namely a RF with 15 trees and a maximum depth of 10 is chosen here. 
The memory size can also be compressed by reducing the dimension of the input images, either by preprocessing the data with principal component analysis (PCA), independent component analysis (ICA) or with an autoencoder. Fig.\ref{fig:4}g shows a standard variational autoencoder schematic, with 128 hidden neurons in the decoder and encoder path and a latent space of 32 elements. We trained the RF with the data set pre-processed with the autoencoder, namely with a feature vector size of 32. Fig. \ref{fig:4}h shows the classification accuracy as function of maximum depth of the trees with and without autoencoder pre-processing, where it is possible to obtain a loss of a few percent in exchange for significant compression. Fig. \ref{fig:4}i shows the memory size using $32\times32$ tiles, for encoding the RF with and without autoencoder, demonstrating a compression factor close to the latent space dimension compared with the original dimension.

\subsection*{Performance evaluation}
\begin{figure}
\centering
\includegraphics[width=0.9\linewidth]{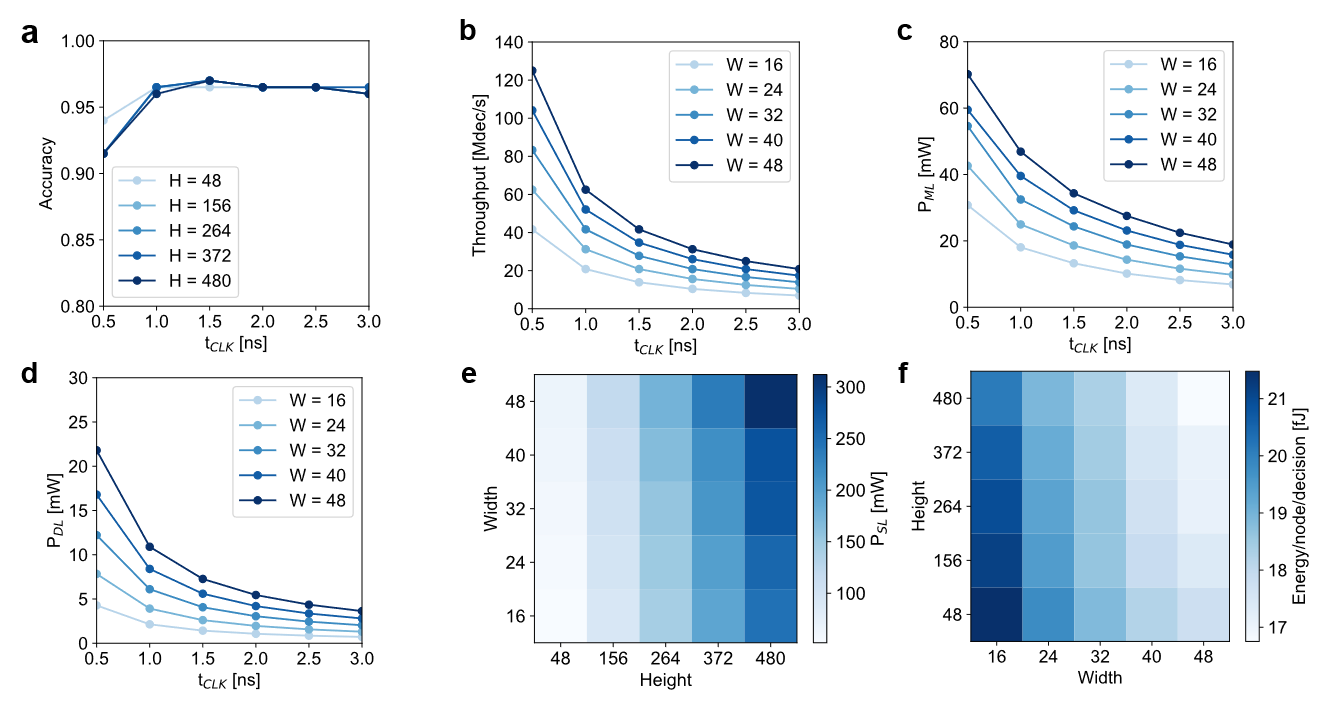}
\caption{Performance evaluation. (a) Classification accuracy as function of $t_{clk}$ for different tile height $H$ at fixed tile width $W=16$. Accuracy does not depend on $H$ as expected, until for nominal operational frequency namely for $t_{clk}>1 ns$, below which mapping in the chosen conductance range is not possible. Throughput (b), ML dynamic power consumption (c) and DL dynamic power consumption (d) as function of $t_{clk}$ for different $W$ at fixed $H=48$. (e) SL static power consumption as function of $H$ and $w$ for $t_{clk}=1 ns$. (f) Energy per DT node per decision as function of $H$ and $w$ for $t_{clk}=1 ns$.}
\label{fig:5}
\end{figure}

To evaluate the power consumption and throughput of the analog CAM system, we considered a full circuit model including ML pre-charge circuit, sense amplifier, a digital to analog converter \cite{li_analog_2020} for charging the DL (Supplementary Information Fig.\ref{sifig:6}) and memristor conductances from the data of Fig.\ref{fig:1}e. To study the design hyperparameters namely tile size $H$ and $W$, and clock frequency $t_{CLK}$ we evaluate accuracy, throughput, power consumption and energy per node per decision, namely the energy spent for assessing threshold in a tree.Fig.\ref{fig:5}a shows accuracy as function of $t_{CLK}$ for different $H$ and a fixed $W=16$. As expected, accuracy does not depend on $H$ but there is a dependence on $t_{CLK}$, as enough time should be given to ML for discharging if the input does not correspond to a match. However, a good accuracy is preserved for $t_{CLK}>1 ns$, which guarantees a high throughput up to $60\times 10^{6}$ Decisions/sec, as shown in Fig.\ref{fig:5}b. Note that the throughput does depend on $W$, in fact arrays $A_i$ are evaluated one by one, thus a smaller tile size $W$ corresponds to a larger number of arrays $A$ and latency. However, with a continuous input flow, operations can be pipelined by for example charging ML of $A_1$ while latching the ML result of $A_0$ and throughput can be highly increased at the cost of power consumption, at fixed energy per decision. Fig.\ref{fig:5}c shows the dynamic power needed for charging and discharging the ML as function of $t_{CLK}$ for different $W$, demonstrating a dependence on $W$ due mostly to tiling. In Fig.\ref{fig:5}d it is reported the dynamic power consumption needed for charging the DL, considering the proposed DAC \cite{li_analog_2020} and an optimum $R_{out}$ (Supplementary Information \ref{sifig:6}). While dynamic power consumption is low, static power consumption due to the voltage divider evaluation of M1-T1 and M2-T3, which is shown in Fig.\ref{fig:5}e as function of $H$ and $W$ is quite important and should be carefully taken into account while choosing the hyperparameters. Finally, Fig.\ref{fig:5}f shows the energy per node per decision as function of $H$ and $W$ at $t_{CLK} = 1 ns$. Taken into account all the dependencies we decide for using tiles of sizes $16\times480$.\\
We finally evaluated our result on a 16 nm technology\cite{li_analog_2020}, by applying a constant field scaling procedure. We considered a maximum clock frequency of 1 GHz, as a practical case \cite{kang_194-njdecision_2018}, and evaluated the performance of our architecture compared with different results from literature \cite{van_essen_accelerating_2012,chen_visual_2012,lee_vocabulary_2015,kang_194-njdecision_2018}. The comparison is shown in Table \ref{tab:1} with analog CAM outperforming existing accelerators in throughput and energy per decision. By using a pipe-lined architecture one can modulate the amount of power spent at the cost of a reduced throughput or vice-versa, in a constant-energy power/accuracy trade off. We envision that such results open the possibility for analog CAM to accelerate different tree-based workloads, including state-of-the-art AI tasks that usually require high energy consumption for training and inference \cite{schrittwieser_mastering_2020}.

\begin{table}
\centering
\begin{tabular}{|l|l|l|l|l|l|}
\hline 
Accelerator                                             & Process   & $f_{clk}$ [Ghz]   & Power [mW]            & Throughput [Dec/s]   & Energy [nJ/dec] \\
\hline
Intel X5560 \cite{van_essen_accelerating_2012}          & 45nm      & 2.8               & 190 $\times 10^3$     & 9.3 $\times 10^3$     & 20.4 $\times 10^6$ \\
\hline
Nvidia Tesla M2050 \cite{van_essen_accelerating_2012}   & 40nm      & 2.8               & 225 $\times 10^3$     & 20.4 $\times 10^3$    & 11 $\times 10^6$ \\
\hline
Xilinx Virtex-6 \cite{van_essen_accelerating_2012}      & 40nm      & 0.079             & 11 $\times 10^3$      & 31.3 $\times 10^3$    & 351 $\times 10^3$ \\
\hline
ASIC \cite{chen_visual_2012}                            & 65nm      & 0.2               & 5.6                   & 30                    & 186.7 $\times 10^3$ \\
\hline
ASIC \cite{lee_vocabulary_2015}                         & 65nm      & 0.25              & 27.6                  & 60                    & 460 $\times 10^3$ \\
\hline
ASIC IMC \cite{kang_194-njdecision_2018}                & 65nm      & 1                 & 7.1                   & 364.4 $\times 10^3$   & 19.4 \\
\hline
This work                                               & 16nm      & 1                 & 3.62                  & 20.83 $\times 10^6$    & 0.17 \\
\hline
This work pipelined                                     & 16nm      & 1                 & 58                    & 333 $\times 10^6$    & 0.17 \\
\hline
\end{tabular}
\caption{\label{tab:1}Comparison of tree-based ML accelerators in literature with our work, which shows up to $10^3$x better throughput and up to $\sim$ 2x less power consumption.}
\end{table}

\section*{Discussion}
In summary, we have proposed a tree-based machine learning accelerator with IMC primitives based on analog CAM, in which by mapping root-to-leaf paths to CAM array rows it is possible to perform rapid inference. A post layout compact model of the analog CAM was designed to assess performance on RF inference as part of a larger CAM-RRAM system implementation. Results at a scaled technology node demonstrate up to $\sim$2x lower power consumption and $\sim10^3$x higher throughput, resulting in $\sim10^2$x reduced energy per decision compared with the state-of-the-art. The high performance offered for this class of machine learning models possessing increased explainability provides a compelling opportunity to use analog CAM-based hardware in critical application areas.

\cleardoublepage
\section*{Methods}
\subsection*{Memristor integration}
The memristors were monolithically integrated on CMOS fabricated in a 180 nm technology node. The integration starts with a removal of silicon nitride and oxide passivation with reactive ion etching (RIE) and a buffered oxide etch (BOE) dip. Chromium and platinum bottom electrodes are then patterned with e-beam lithography and metal lift-off process, followed by reactive sputtered 4.5 nm tantalum oxide as switching layer. The device stack is finalized by e-beam lithography patterning of sputtered tantalum and platinum metal as top electrodes.
\subsection*{Analog CAM circuit simulation}
6T2M analog CAM cell and small arrays were designed and simulated in Cadence Virtuoso Custtom IC design environment, and the simulation result post-processed with HP-SPICE. The simulations utilize the TSMC 180 nm and 16 nm library and the designs follow the corresponding rules. A custom python script generates the netlist for analog CAM arrays with different numbers of rows and columns and arbitrary configured memristor conductance and input voltages.
\subsection*{Analog CAM model}
Analog CAM model was implemented in Python environment by fitting outputs of circuit simulation with simplified physical laws and behavioral equation. While mostly in subthreshold regime, input transistor T1(T3) was modeled both in subthreshold and ohmic conduction regime. The first obey to the simplified MOS model:
\begin{equation}
    I_{D1}=I_{D0}exp\left(\frac{V_{DL}}{\alpha}\right)
\end{equation}
with $I_{D0}$ and $\alpha$ fitting parameters (for details see Supplementary Information \ref{sitab:1}). Given that T1 drain-source voltage (corresponding to the voltage divider) $V_{div}=V_{G,T2}-V_{SL,lo}$ is typically low, or $V_{div}<V_{GS}-V_T$ with $V_{GS}=V_{DL}-V_{SL,lo}$ and $V_T$ threshold of T1, in the region of interest, we assume that it can only be either in sub threshold or ohmic (linear) region, whose current obeys to the law:
\begin{equation}
    I_{D1}=k_1(V_{GS}-V_T)
\end{equation}
with $k_1$ fitting parameter corresponding to physical and electrical properties. Once the voltage divider has been computed, the ML discharge current can be computed as:
\begin{equation}
    I_{D2}=k_2(V_{div}-V_T)^2
\end{equation}
Being the output transistor T2 (T6) typicall biased in the saturation region, at least in the initial discharging phase. Finally, the inverter was modeled as a sigmoid for simplicity and fast calculation:
\begin{equation}
    V_{G.T6}=\frac{-0.8}{1+exp(-\beta(V_{div}+\gamma))}+0.8.
\end{equation}
Parastic parameters were extracted from the post-layout simulation and corresponds to a resistance connecting each cell namely $R_{ML}=R_{DL}= r = 1.4\Omega$ and a parasitic capacitance of $C_{ML}=C_{DL} = c = 1.9 fF$. Pre charge block and sense amplifier were assumed as a parasitic capacitance which were extracted from the post layout simulation as $C_{PC}=40.95 fF$ and $C_{SA}=50fF$ respectively.
\subsection*{Models training}
All tree-based model were trained in a Python environment with sklearn module. To match the benchmark, we trained RF with same 8 class and training/testing set as in literature\cite{kang_194-njdecision_2018}, thus we used 2300 training and 200 testing images. However, while the reference RF was trained with 64 trees and a maximum depth of 6, we optimized the hyperparameters namely maximum depth and number of trees reaching an accuracy of 96.5 \% when deployed to analog CAM, overcoming the given accuracy of 94\%. 
\subsection*{Power consumption calculation}
Power consumption calculation of the pipelined architecture was divided in three parts, namely 
\begin{itemize}
    \item static power consumption flowing into the voltage divider
    \begin{equation}
    P_{static}=V_{sl,hi}I_{D0}
    \end{equation}
    \item dynamic power consumption to charge the DL
    \begin{equation}
    P_{DL}=\frac{V_{DD}^2WN}{R}
    \end{equation}
    with $W$ tile width, $N$ number of tiles and $R$ output resistance of the DAC (Supplementary Information)
    \item dynamic power consumption to charge and discharge the ML
    \begin{equation}
    P_{ML}=\frac{1}{2t_{CLK}}(C_{ML}V_{ML0})^2HN + \sum_{j=0}^{i=N}\sum_{i=0}^{i=H}\frac{1}{2t_{CLK}}(C_{ML}(V_{ML0}-V_{ML,i,j}))^2
    \end{equation}
    with $V_{ML0}$ initial voltage of the ML, which can be set from the pre charge block, $H$ tile height and $V_{ML,i,j}$ ML voltage of row $i$ of tile $j$ at $t=t_{CLK}$. The first term corresponds of the charging energy and the second to the discharging in each cell.
\end{itemize}

\cleardoublepage
\bibliography{references.bib}





\section*{Author contributions statement}
G.P, C.G, C.L and J.P.S. conceived the tree mapping procedure, G.P. design the compact model, C.L., X.S., R.M. conducted experiment, G.P. and C.L. analyzed the data, G.P. and S.S. designed the machine learning models, G.P.,C.G. and J.P.S conceived the compressing procedure and architecture. All authors reviewed the manuscript. J.P.S. supervised the research.







\end{document}


\flushbottom
\maketitle

\begin{figure}
\centering
\includegraphics[width=0.7\linewidth]{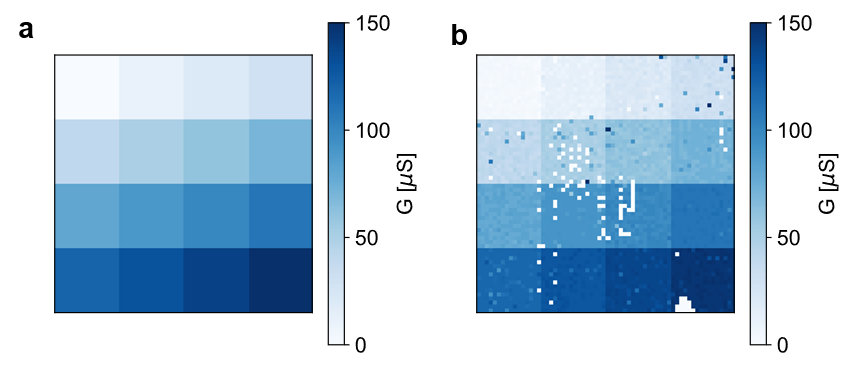}
\caption{Memristor programming. (a) Target $64\times64$ matrix divided in 16 different box, each comprising 256 elements of different values. (b) Results of the measured programmed matrix. An integrated circuit \cite{li_cmos-integrated_2020} comprising 3 $64\times64$ memristor arrays, sensing and routing circuitry was used to study the noise in programming each level. Some devices were stuck in a low (white) or high conductance (strong blue) that can not be recovered. For this reason we considered only the best 128 out of 256 devices for the cumulative distribution plots. }
\label{sifig:1}
\end{figure}

\clearpage

\begin{figure}
\centering
\includegraphics[width=0.9\linewidth]{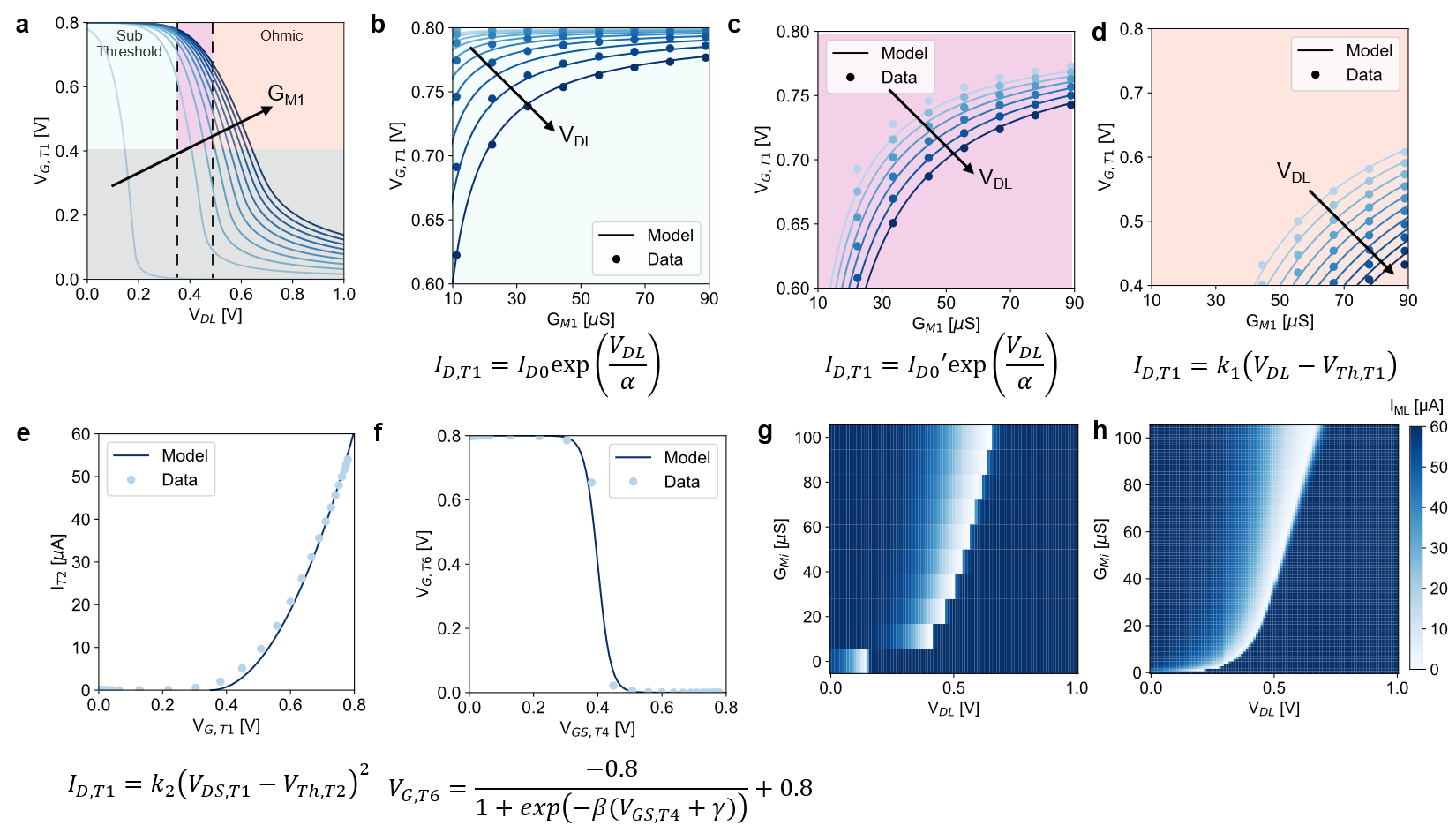}
\caption{Analog CAM model details. (a) Circuit simulations results $V_{G,T1}$ as function of $V_{DL}$ for different programmed $G_{M1}$ evidencing two main working region namely sub threshold for $V_{DL}<0.3$ (light blue) and ohmic (or linear) for $V_{DL}$>0.5 (orange). We also considered an intermediate subthreshold region for $0.3<V_{DL}<0.5$ (purple) with subtreshold conduction but slightly different fitting parameters. Circuit simulation results and model calculation of $V_{G,T1}$ as function of $G_{M1}$ in the subthreshold regime (b), intermediate regime (b) and ohmic regime (c). Equations for fitting curves are shown below. (e) Circuit simulation results and calcualtion of current flowing from ML through T2 as function of $V_{G,T1}$, in this case a saturation regime is considered. Equation for fititng is shown below. (f) Circuit simulation results and behavioral model of the inverter, with the model equation below. (g) Circuit simulations results and (h) model calculation of the overall discharge currant as function of $V_{DL}$ and $G_{Mi}$ assuming the same conudctance programmed in both memristor of a analog CAM cell. Model fitting parameters are shown in Supplementary Table \ref{sitab:1}}
\label{sifig:2}
\end{figure}

\clearpage

\begin{table}
\centering
\begin{tabular}{|l|l|}
\hline 
Constant                & Value \\
\hline
$I_{D0}$                & 50 nA \\
\hline
$\alpha$                & 80 mV \\
\hline
$I_{D0}'$               & 45 nA \\
\hline  
$k_{1}$                 & 160 $\mu$A/V \\
\hline
$V_{th,T1}$             & 405 mV \\
\hline
$k_{2}$                 & 300 $\mu$A/V$^2$ \\
\hline
$V_{th,T2}$             & 350 mV \\
\hline
$\beta$                 & 50 V$^{-1}$ \\
\hline
$\gamma$                & -0.4 V \\
\hline
\end{tabular}
\caption{\label{sitab:1} Analog CAM model fitting parameters}
\end{table}

\clearpage

\begin{figure}
\centering
\includegraphics[width=0.3\linewidth]{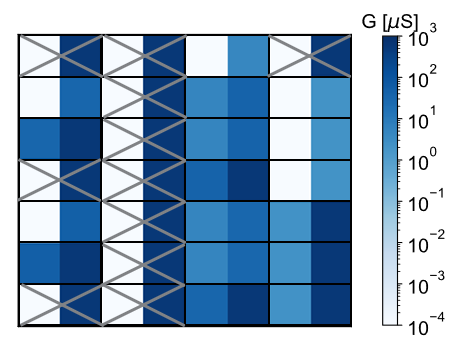}
\caption{Programmed conductance in a analog CAM arrays for the classification of Iris dataset, 'X' corresponds to 'don't care' namely memristor programmed in the wider possible range with M1 = HRS and M2 = LRS}
\label{sifig:3}
\end{figure}

\clearpage

\begin{figure}
\centering
\includegraphics[width=0.3\linewidth]{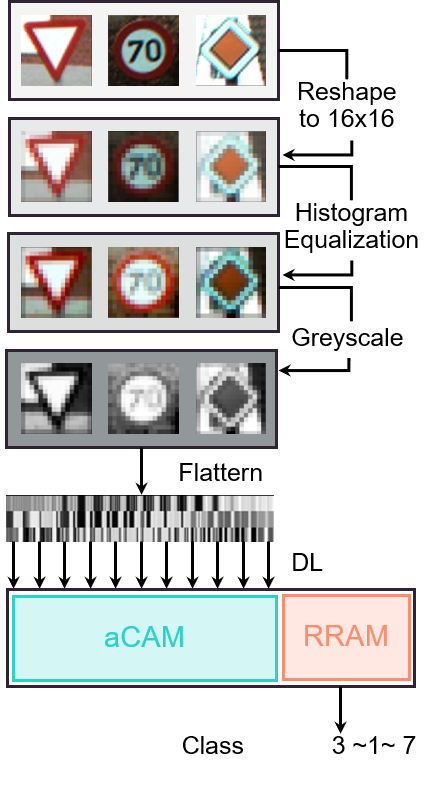}
\caption{Random forest training pre processing. The same procedure of the reference benchmark\cite{kang_194-njdecision_2018} was applied. KUL traffic sing data set consist of $32\times32$ 256 levels RGB image that were reshaped to $16\times16$. After that an histogram equalization technique was applied and finally the image was converted to grey scale. After flattening to a $1\times256$ array, the image is applied to analog CAM DL and the matched class activate the corresponding memristors on the RAM array for majority voting.}
\label{sifig:4}
\end{figure}

\clearpage

\begin{figure}
\centering
\includegraphics[width=0.8\linewidth]{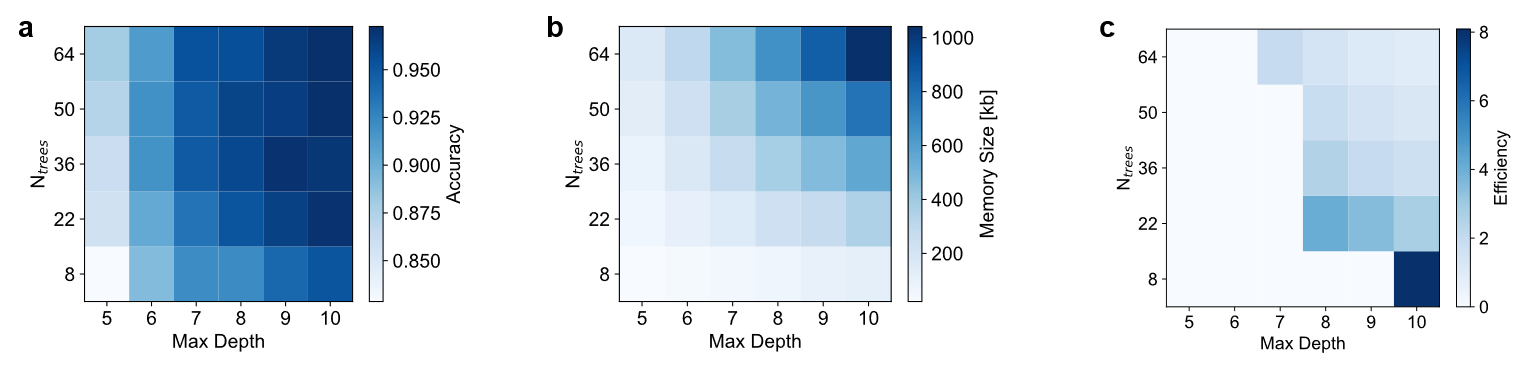}
\caption{Random Forest hyperparameters tuning. (a) Accuracy as function of number of tree $N_{trees}$ and maximum depth of each tree. (b) Memory size with $32\times32$ tiles as function of $N_{trees}$ and maximum depth of each tree. (c) Efficiency namely $Efficiency = Accuracy/Memory Size$ as function of $N_{trees}$ and maximum depth of each tree. Points with accuracy < 0.96 were excluded to be sure of matching the reference 0.94 accuracy \cite{kang_194-njdecision_2018} once all non idealities are included.}
\label{sifig:5}
\end{figure}

\clearpage

\begin{figure}
\centering
\includegraphics[width=0.8\linewidth]{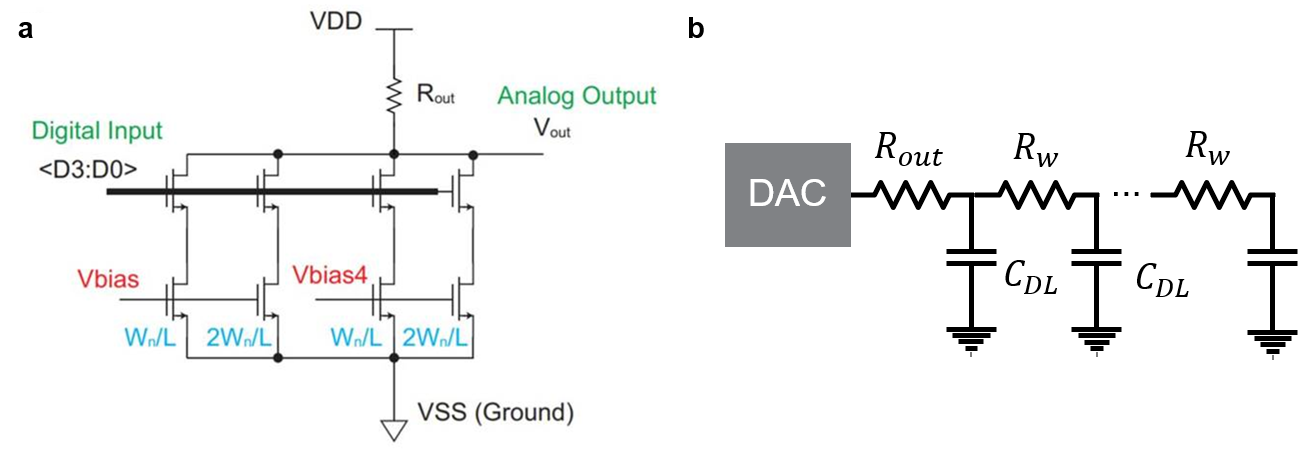}
\caption{Digital to analog converter (DAC) design and DL power consumption calculation. (a) Schematic of the current steering DAC considered for this work \cite{li_analog_2020}. $R_{out}$ was optimized by considering the clock time and parasitic resistance and conductance due to wires. (b) Schematic of a DL circuit where each analog CAM has a parasitic capacitance $C_{DL}$ and analog CAM cells are connected with each other with a wire of resistance $R_{w}$. Elmore's theorem was considered to compute the propagation delay namely $\tau=\sum_{i=0}^{H}(R_{out}+iR_w)C_{DL} = C_{DL}\left[R_{out}H+R_w\frac{H\left(H-1\right)}{2}\right]$, with $H$ height of the analog CAM tile, or number of analog CAM cells in a column.} 
\label{sifig:6}
\end{figure}

\clearpage
\bibliography{references.bib}